\documentclass[fleqn,usenatbib]{mnras}

\usepackage{newtxtext,newtxmath}
\usepackage[T1]{fontenc}
\usepackage{ae,aecompl}

\usepackage{graphicx}	% Including figure files
\usepackage{amsmath}	% Advanced maths commands
\usepackage{amssymb}	% Extra maths symbols
\usepackage{bm}
\usepackage{color}

\title[Magnetic Energy Spectrum in Tycho's SNR]{Discovery of Kolmogorov-like Magnetic Energy Spectrum
in Tycho's Supernova Remnant by Two-point Correlations of Synchrotron Intensity}

\author[J. Shimoda et al.]{
Jiro Shimoda,$^{1,2}$\thanks{E-mail: s-jiro@phys.aoyama.ac.jp, j-shimoda@astr.tohoku.ac.jp (JS)}
Takuya Akahori$^{3}$
A. Lazarian,$^{4}$
Tsuyoshi Inoue,$^{5}$
and Yutaka Fujita$^{6}$
\\
$^{1}$Depertment of Physics and Mathematics, Aoyama-Gakuin University, Sagamihara, Kanagawa 252-5258, Japan\\
$^{2}$Frontier Research Institute for Interdisciplinary Sciences, Tohoku University, Sendai 980-8578, Japan\\
$^{3}$Mizusawa VLBI Observatory, National Astronomical Observatory of Japan, 2-12 Hoshigaoka, Mizusawa, Oshu, Iwate 023-0861, Japan\\
$^{4}$Department of Astronomy, University of Wisconsin, 475 North Charter Street, Madison, WI 53706, USA\\
$^{5}$Department of Physics, Graduate School of Science, Nagoya University, Furo-cho, Chikkusa-ku, Nagoya 464-8602, Japan\\
$^{6}$Departmemt of Earth and Space Science, Gradate School of Science, Osaka University, Toyonaka, Osaka 560-0043, Japan
}

\date{Accepted XXX. Received YYY; in original form ZZZ}

\pubyear{2017}

\begin{document}
\label{firstpage}
\pagerange{\pageref{firstpage}--\pageref{lastpage}}
\maketitle

\begin{abstract}
The spectral slope of the magnetic energy in supernova remnants (SNRs) can be obtained by
analysis of spatial two-point correlation functions of synchrotron intensities.
This method has been originally developed for the analysis of magnetic field structure
in diffuse interstellar medium and applied when the geometry of the emission region is simple and known.
In this paper, applying this correlation analysis to Tycho's SNR,
for which the synchrotron emission region is known to be a spherical shell,
we find that the magnetic energy spectrum shows the Kolmogorov-like scaling.
Our results can be explained by turbulence developed downstream of the shock front via
Richtmyer-Meshkov instability or
an amplification of upstream magnetic field induced by cosmic rays.
They could be discriminated by future observations with
a sub arcsecond resolution such as Square Kilometer Array.
\end{abstract}


%%%%%%%%%%%%%%%%%%%%%%%%%%%%%%%%%%%%%%%%%%%%%%%%%%

%%%%%%%%%%%%%%%%% BODY OF PAPER %%%%%%%%%%%%%%%%%%

\section{Introduction}
Supernova remnant (SNR) shock waves are believed to be an accelerator of Galactic cosmic-rays (CRs)
with energies at least up to $10^{15.5}~{\rm eV}$ (called ``knee energy'').
However, there is no firm evidence that the SNRs are accelerators of the knee-energy CRs.
In the standard scenario, the CR particles are accelerated 
through the ``diffusive shock acceleration'' (DSA) mechanism~\citep[e.g.][]{blandford78,bell78}
which is accompanied by simultaneous generation of magnetic-field disturbances
at the vicinity of the shocks~\citep[e.g.][]{bell04}.
In the DSA mechanism, CR particles are scattered
through interactions with the field disturbances
to go back and forth between upstream and downstream of the shock,
suffering the shock heating repeatedly and increasing their energy.
If the mean free path of the accelerated particles is sufficiently larger than
a radius of the SNR,
they escape from the SNR shock and the acceleration is finished.
Thus, the maximum energy of the accelerated particles depends on their diffusion coefficient.
If we consider a monochromatic field disturbance with scale length $l$,
a CR particle with gyroradius $r_g\simeq l$ interacts resonantly with
the monochromatic field disturbance
resulting in a pitch-angle scattering~\citep[][]{jokipii66}.
The strength of the scattering depends on an energy density of the magnetic field disturbance.
Because the field disturbance
usually has a continuous energy spectrum due to a turbulent cascade,
the CR particles with different energies resonate with the field disturbances at different scale lengths.
Hence,
the diffusion coefficient
and the maximum energy are related to the spectrum.
For the field disturbances with an energy spectrum proportional to $l^m$,
the diffusion coefficient of the particles with energy $E$ can be written as
%
%%%%%%%%%
\begin{eqnarray}
\kappa(E)\sim\tilde\kappa(E_0)\left(\frac{E}{E_0}\right)^{1-m},
\label{kappa m}
\end{eqnarray}
%%%%%%%%%
%
where $E_0$ is the energy of the CR particle giving the representative diffusion coefficient
$\tilde\kappa(E_0)=cr_g(E_0)/3$~\citep[see, e.g.][]{blandford87,parizot06}.
$c$ is the speed of light.
\citet{parizot06} evaluated the maximum energy of CR protons as a function of the spectral slope $m$
for a number of SNRs
(Cas~A, Kepler, Tycho, SN~1006 and G347.3-0.5) by deriving the magnetic-field strength
from the thickness of non-thermal X-ray filaments.
For Tycho's SNR, if $m\sim2/3$ (corresponding to Kolmogorov-like turbulence),
the maximum energy of CR protons reaches around the knee energy.
\footnote{$\gamma$-ray emissions from Tycho's~SNR may originate in high energy CR protons
with an energy at least $\sim10^{14}~{\rm eV}$
but there is no consensus on the maximum energy of CR protons because it depends on emission models
~\citep[see][]{archambault17}.}
While the slope of the magnetic energy spectrum should be related to the maximum energy of CRs,
it has not been well determined observationally.
Moreover, there is a controversy over the shape of the magnetic energy spectrum
realized in the SNRs
such as a single power-law~\citep[e.g.][]{gs95,cho00b},
a broken power-law~\citep[e.g.][]{lazarian99,cho00a,cho03,brandenburg05,lazarian06,beresnyak09,inoue12,xu16,xu17}
and a spectrum containing several discrete peaks~\citep[][]{vladimirov09}.
The uncertainty of the spectrum prevents us from determining the maximum energy based on
the standard scenario.
Therefore, it is important to obtain the spectral slope observationally to reveal
acceleration process of CRs at SNRs.
\par
It is widely recognised for interstellar medium (ISM)
that the two-point correlation function of synchrotron intensities reflects
statistical nature of turbulent magnetic field including the spectral slope of magnetic energy
~\citep[e.g.][see also \citet{akahori18} for a review]{getmantsev59,chepurnov98,cho10,lazarian12,lazarian16}.
This correlation analysis, however, suffers from the uncertainty of the geometry of the emission region
because of the projection effect.
In other words, the correlation function reflects not only the structure of the magnetic field
but also the geometry of the emission region.
Therefore, the geometry must be determined separately to obtain the spectral slope
of magnetic energy.
Fortunately, the emission region of young SNRs is often a spherical shell~\citep[e.g.][]{dickel91,reynoso13}.
Thus, if we select arbitrary points on a concentric circle of an SNR image,
the depth of the emission region along the line of sight through the points is constant.
This means that we can exclude the geometrical effect from the correlation function.
As a result, the derived correlation function depends only on the spectral slope of magnetic energy.
\par
In this paper, we explore the spectral slope of magnetic energy in Tycho's SNR
by applying the correlation analysis.
This paper is constructed as follows.
In section~\ref{sec analysis}, we briefly explain our analysing method and
the application for Tycho's SNR.
The results are shown in section~\ref{sec results},
and we discuss the origin of magnetic field structure in Tycho's SNR in section~\ref{sec discussion}.

%%%%%%%%%%%%%%%%%%%%%%%%%%%%%%%%%%%%%%%%%%%%%% Section 2

%%%%%%%%%%%%%%%%%%%%%%%%%%%%%%%%%%%%%%%%%%%%%% Section 3
\section{Analysis of Magnetic Field Correlation}
\label{sec analysis}
To extract magnetic field correlation,
we consider the second-order correlation function of
the synchrotron intensity per frequency $I_\nu$
on the circle $S$ centred on the SNR centre.
The function is given by
%
%%%%%%
\begin{eqnarray}
C^{(2)}_{I_\nu,S}(\bm{\lambda}) &=& \frac{\int_{S} I_\nu(\bm{X})I_\nu(\bm{X'}) d^2\bm{X}}
{\int_{S}d^2\bm{X}} \nonumber \\
& \equiv & \langle I_\nu(\bm{X})I_\nu(\bm{X}+\bm{\lambda}) \rangle_{\bm{X},S},
\label{correlation}
\end{eqnarray}
%%%%%%
%
%where
where $\bm{X}=(x,y)$ is the two-dimensional sky position and
$\bm{\lambda}=\bm{X'}-\bm{X}$ is the position vector of two separated
sky positions $\bm{X}$ and $\bm{X'}=\bm{X}+\bm{\lambda}$
 (see Appendix~\ref{sec measurements method} for detail).
Here we select $\bm{X}$ and $\bm{X}'=\bm{X}+\bm{\lambda}$ from the region
%
%%%%%%
\begin{equation}
S(\bm{X},R) = \left\{\bm{X}~\big|~(R-\delta)^2\le f(x,y)\le (R+\delta)^2
\right\},\label{arc}
\end{equation}
%%%%%%
%
with $f(x,y)=(x-x_c)^2+(y-y_c)^2$,
where $(x_c,y_c)$ is the centre of the SNR,
$R$ is the radius of the circle we are interested in and
$\delta\ll R$ is the width.
Note that although the correlation function $C^{(2)}_{I_\nu,S}(\bm{\lambda})$ is defined in a two-dimensional space,
it is mostly represented as one-dimensional function owing to the condition $\delta \ll R$,
that is, the domain of definition of $C^{(2)}_{I_\nu,S}(\bm{\lambda})$
is much elongated in the azimuthal direction.
For Kolmogorov-like turbulent field, the one-dimensional correlation function
shows the scaling relation of $\lambda^{2/3}$~\citep[e.g.][]{kolmogorov41}.
\par
\begin{figure}
\center
\includegraphics[scale=0.35]{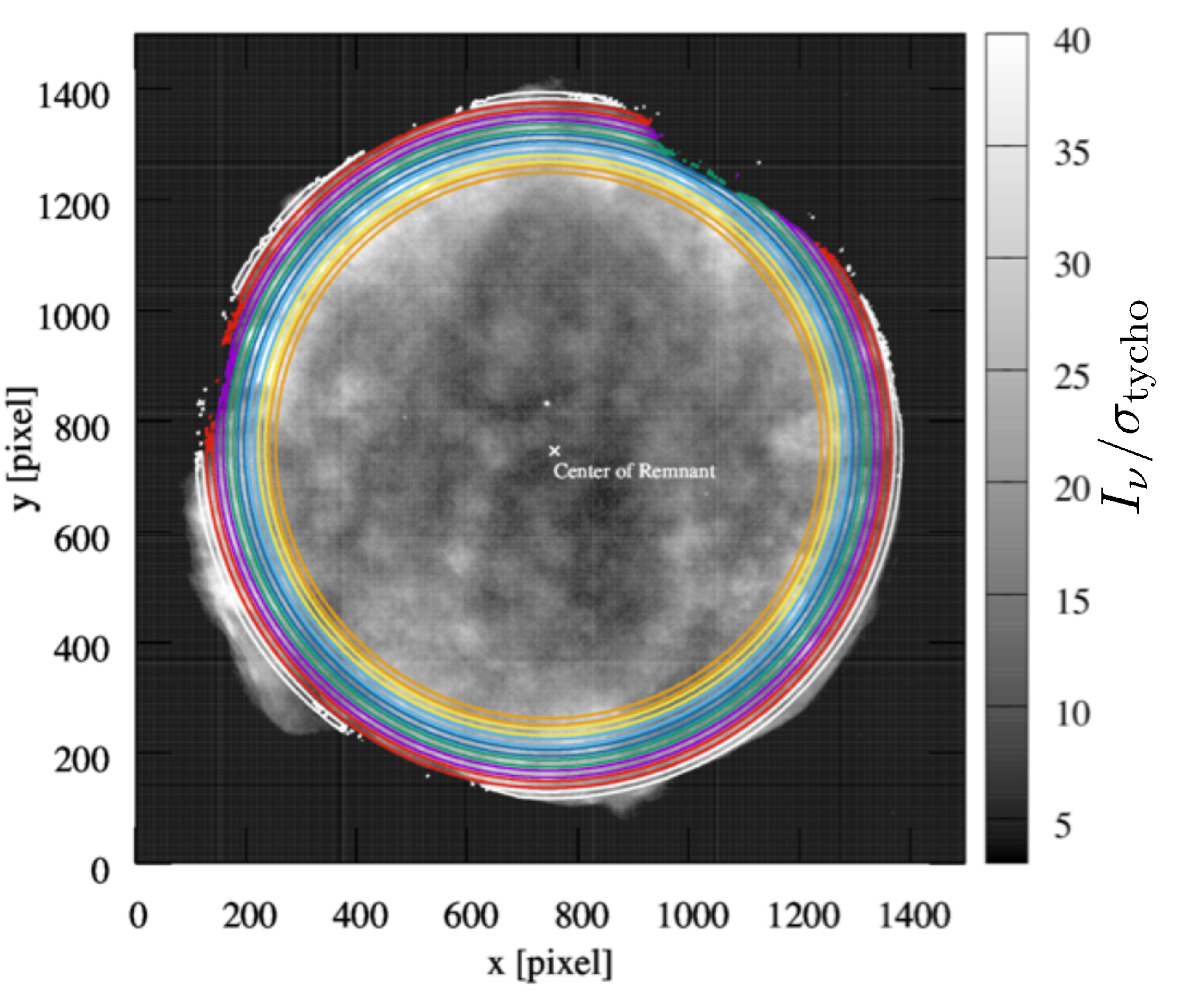}
\caption{
Radio synchrotron ($I_\nu$) images of Tycho's SNR.
$\sigma_{\rm tycho}=5.3\times10^{-5}~{\rm Jy~beam^{-1}}$
is the image noise level.
$x$ and $y$ axes are in units of the number of pixels
(one pixel size is $0.4$ arcsec).
The origin of the coordinates (J2000) for Tycho's SNR is
${\rm (R.A.,Dec.)}=(0^{\rm h}25^{\rm m}19^{\rm s}.1,+64^\circ08'23''.0)$.
The regions enclosed by coloured lines
(white, red, purple, green, blue, light blue, yellow and orange)
indicate $I_{\nu}\ge3\sigma_{\rm Tycho}$ at each concentric circles
($R=1.00~R_{\rm SNR}$,
$0.97~R_{\rm SNR}$,
$0.94~R_{\rm SNR}$,
$0.91~R_{\rm SNR}$,
$0.88~R_{\rm SNR}$,
$0.85~R_{\rm SNR}$,
$0.82~R_{\rm SNR}$ and
$0.79~R_{\rm SNR}$).
}
\label{tycho image}
\end{figure}

We study Tycho's SNR using the correlation function.
We analyse a $1.4~{\rm GHz}$ image, which is published in \citet{williams16},
obtained by Very Large Array: project VLA/13A-426 (PI J. W. Hewitt).
Figure~\ref{tycho image} shows the image.
The synthesized beam size (angular resolution) is $1.92~{\rm arcsec}$ and the image pixel size is $0.4~{\rm arcsec}$.
The image noise level is $\sigma_{\rm Tycho}=5.3\times10^{-5}~{\rm Jy~beam^{-1}}$.
Supposing that the distance
to Tycho's SNR is $d=4~{\rm kpc}$~\citep[][]{hayato10,katsuda10},
$1~{\rm arcsec}\approx0.02~{\rm pc}(d/4~{\rm kpc})$.
We calculate the intensity centroid from pixels with $I_{\nu}\ge3\sigma_{\rm Tycho}$ at the rim of SNR,
and set the centre of concentric circles at the centroid:
${\rm (R.A.,Dec.)}=(0^{\rm h}25^{\rm m}19^{\rm s}.1,+64^\circ08'23''.0)$.
It mostly agrees with the geometrical centre derived from the X-ray image~\citep{ruiz04}:
 ${\rm (R.A.,Dec.)}=(0^{\rm h}25^{\rm m}19^{\rm s}.9,+64^\circ08'18''.2)$.
We analyse eight concentric circles with radii,
$R=1.00R_{\rm SNR}$ (white),
$0.97R_{\rm SNR}$ (red),
$0.94R_{\rm SNR}$ (purple),
$0.91R_{\rm SNR}$ (green),
$0.88R_{\rm SNR}$ (blue),
$0.85R_{\rm SNR}$ (light blue)
$0.82R_{\rm SNR}$ (yellow)
and $0.79R_{\rm SNR}$ (orange),
where $R_{\rm SNR} = 632~{\rm pixels} \approx 253~{\rm arcsec} \approx 5~{\rm pc}$ is the SNR radius.
The width of each circle is set to be $\delta=0.0091R_{\rm SNR}$.
\par
Errors on the correlation function are evaluated from the lower and upper limits of $C^{(2)}_{I_\nu,S}$.
There are pixels having weak signals of $I_\nu<3\sigma_{\rm Tycho}$.
We regard such weak signals as noise.
Meanwhile,
if we assign a pseudo-signal of $3\sigma_{\rm Tycho}$ to those pixels,
we obtain the maximum value of the correlation function.
We regard it as the upper limit.
Similarly, if we assign zero to those pixels,
we obtain the minimum value as the lower limit.
We regard these limits as the errors of $C^{(2)}_{I_{\nu},S}$.
%
%%%%%%
%%%%%%
%

\section{Results}
\label{sec results}
%
%%%%%%
\begin{figure}
\center
\includegraphics[scale=0.35]{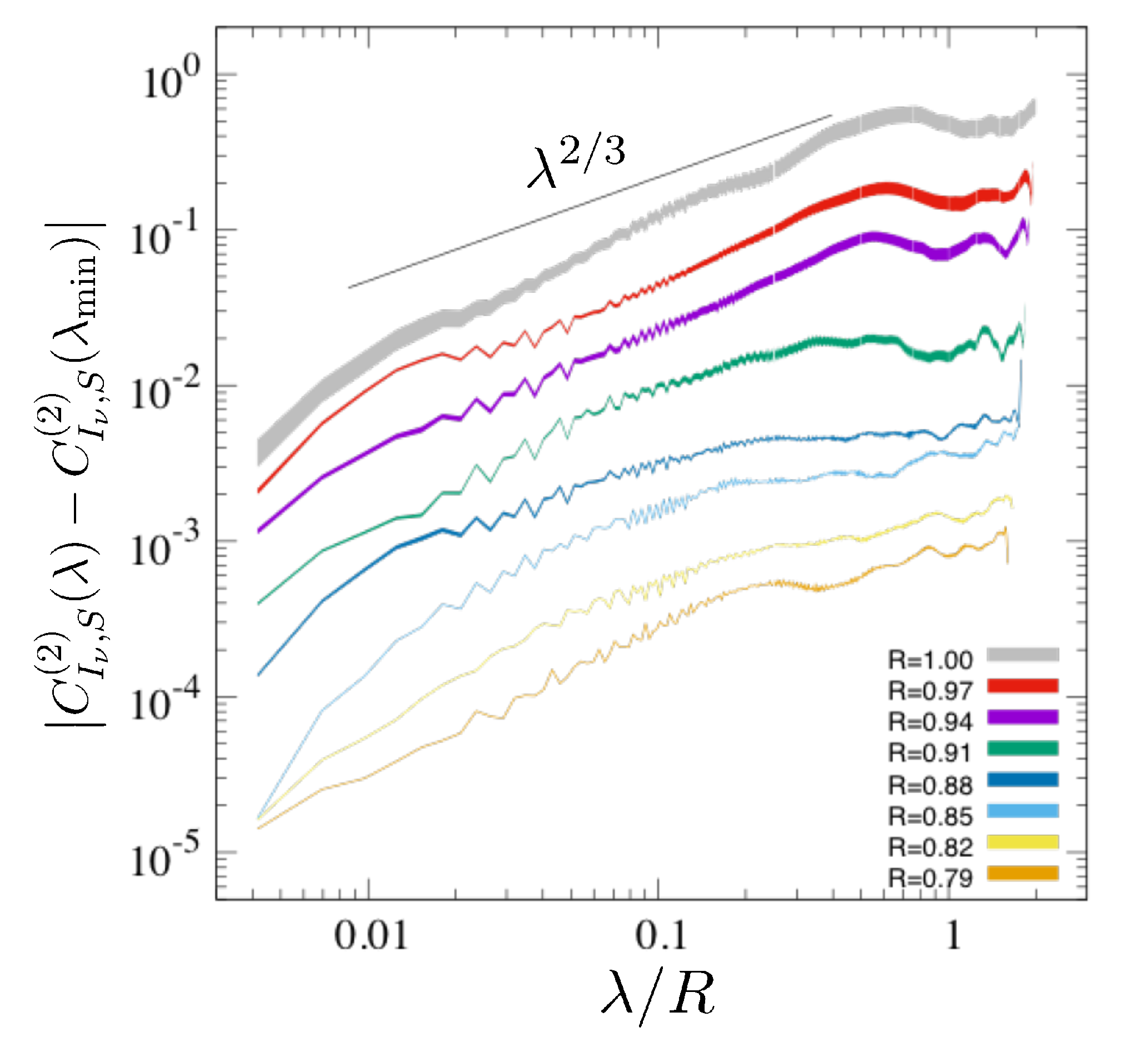}
\caption{
Second-order correlation functions of the observed synchrotron intensities per frequency $I_\nu$,
which imply the spectra of magnetic energy in Tycho's SNR.
To clarify the fluctuating component of the correlation,
we display $\big|C^{(2)}_{I_\nu,S}(\lambda)-C^{(2)}_{I_\nu,S}(\lambda_{\rm min})\big|$.
The solid belts represent
the normalized correlation functions evaluated along the concentric circles with radii of
$R=1.00~R_{\rm SNR}$,
$0.97~R_{\rm SNR}$,
$0.94~R_{\rm SNR}$,
$0.91~R_{\rm SNR}$,
$0.88~R_{\rm SNR}$,
$0.85~R_{\rm SNR}$,
$0.82~R_{\rm SNR}$ and
$0.79~R_{\rm SNR}$ from top to bottom, respectively.
}
\label{tycho circle}
\end{figure}
%%%%%%
%
Figure~\ref{tycho circle} shows the second-order correlation functions $C^{(2)}_{I_\nu,S}$.
The line width indicates the errors evaluated from the upper and lower limits of $C^{(2)}_{I_\nu,S}$.
In order to clarify the correlation of fluctuating component,
we display $\big|C^{(2)}_{I_\nu,S}(\lambda)-C^{(2)}_{I_\nu,S}(\lambda_{\rm min})\big|$,
where $\lambda_{\rm min}\approx1.4$~arcsec is the minimum separation distance between $\bm{X}$ and $\bm{X}'$.
\par
For the outer circles with radii $R\ge0.90$,
$C^{(2)}_{I_\nu,S}$ follows a power law and has a positive slope
at relatively small scales ($\lambda/R \la 0.5$),
i.e. larger-scale magnetic field disturbances are predominant.
The slope is close to the Kolmogorov scaling $\lambda^{2/3}$.
Such a single power-law with the Kolmogorov scaling is predicted by \citet{gs95}~(henceforth, GS95) as
the developed magnetohydrodynamics (MHD) turbulence.
On the other hand, for the inner circles with radii $R\le0.85$,
$C^{(2)}_{I_\nu,S}$ are somewhat flatter than the outer ones at relatively large scales $\lambda/R\ga0.2$,
although a Kolmogorov-like scaling is implied in the small scales $\lambda/R\la0.2$.
It may indicate that the nature of
field disturbances varies at $R\la 0.9~R_{\rm SNR}$.
Actually,
by the combination of measurements of X-ray imaging and spectroscopy,
\citet{warren05}
found that the contact discontinuity,
at which Rayleigh-Taylor instability (RTI) works,
is located at $R\approx0.9~R_{\rm SNR}$.

%%%%%%%%%%%%%%%%%%%%%%%%%%%%%%%%%%%%%%%%%%%%%% Section 4
\section{Discussion}
\label{sec discussion}
The correlation functions for $R\ga 0.9~R_{\rm SNR}$ imply the developed GS95 turbulence that is trans-Alfv\'enic.
If this is the case, the velocity dispersion of the largest eddy, $u_{\rm inj}$,
should be close to the Alfv\'en velocity, $C_{\rm A}$,
and it gives the Alfv\'en Mach number of turbulence, $M_{\rm A,turb}\equiv u_{\rm inj}/C_{\rm A}\approx 1$.
For comparison, if we consider the SNR shock velocity~
\citep[$\simeq 5000~{\rm km~s^{-1}}$,~e.g.][]{williams16}
and the magnetic field strength in the ISM \citep[$\sim 3~{\rm \mu G}$,][]{myers78,beck01},
we obtain a high Alfv\'en Mach number of $\sim500$ for the shock.
Thus, in comparison with the shock, our results imply
smaller gas velocity and/or larger magnetic-field strength at $R\ga 0.9~R_{\rm SNR}$.
The non-thermal X-ray filaments with the thickness $\sim0.01~R_{\rm SNR}$ seen in Tycho's SNR imply
a significant cooling of high energy CR electrons,
suggesting the presence of a strong (likely amplified) magnetic-field~\citep[e.g.][]{bamba05}.
\par
We discuss how the condition $M_{\rm A,turb}\simeq1$ is satisfied at the vicinity of the shock.
Multidimensional MHD simulations \citep[e.g.][]{gj07,inoue09,inoue10,inoue12,inoue13} showed that
the SNR shock is rippled owing to the interaction with
density fluctuations pre-existing in the ISM~\citep[e.g.][]{armstrong95}.
Because of the shock rippling,
the velocity component tangential to the shock surface is generated downstream,
yielding the velocity dispersion just behind
the shock~\citep[e.g.][]{mw68,mahesh97,shimoda15}.
Using three-dimensional MHD simulations,
\citet{inoue13} showed that the strength of the downstream velocity dispersion~$\Delta u$
can be expressed by using growth velocity of the Richtmyer-Meshkov instability (RMI)~$u_{\rm RMI}$:
%
%%%%%%%%%
\begin{eqnarray}
\Delta u\simeq u_{\rm RMI}\simeq A \langle u_{\rm sh}\rangle,
\label{urmi}
\end{eqnarray}
%%%%%%%%%
%
where $A=(\Delta\rho/\langle\rho\rangle)/(1+\Delta\rho/\langle\rho\rangle)$ is the Atwood number
and $\langle u_{\rm sh}\rangle$ is the mean shock velocity.
$\Delta\rho$ and $\langle\rho\rangle$ are the dispersion of upstream density fluctuation and the mean upstream density, respectively.
They assumed a weak magnetic field in the upstream region,
$M_{\rm A}=\langle u_{\rm sh}\rangle/C_{\rm A,1}\simeq100$,
where $C_{\rm A,1}$ is the upstream Alfv\'en speed,
and obtained super-Alfv\'enic turbulence (i.e. $M_{\rm A,turb}>1$) behind the shock (though the result depends on $\Delta\rho$, see \cite{inoue13} for details).
The downstream magnetic field is amplified by the turbulent dynamo process induced by the RMI-driven super-Alfv\'enic turbulence.
The amplified field is able to explain the orientations of observed magnetic fields
in young SNRs~\citep[e.g.][]{dickel76,dickel91,reynolds93,delaney02,reynoso13}.
\par
The field amplification becomes significant at a distance
%
%%%%%%%%%
\begin{eqnarray}
d_{\rm RMI}\simeq l_{\Delta \rho}/(r_{c}A),
\label{lRMI}
\end{eqnarray}
%%%%%%%%%
%
from the shock front~\citep[e.g.][]{richtmyer60,sano12,inoue13},
where $l_{\Delta\rho}$ is the scale length of upstream density fluctuations
and $r_c$ is the shock compression ratio.
\citet{williams13} examined
the ambient density of Tycho's~SNR from infrared dust emissions
and found
order-of-magnitude variations in density at the scale length of
$l_{\Delta\rho}\sim R_{\rm SNR}$.
Such density variations are also inferred by observations of the expansion rate of Tycho's SNR
~\citep[][]{williams16}.
Their results imply
$A\approx1$ on the scale $l_{\Delta\rho}\sim R_{\rm SNR}$,
giving $d_{\rm RMI}\sim0.2~R_{\rm SNR}(l_{\Delta\rho}/R_{\rm SNR})(r_c/4)^{-1}$.
Thus, the magnetic field amplification through
the turbulent dynamo induced by the RMI-driven turbulence
may explain the condition $M_{\rm A,turb}\simeq1$
at $R\sim0.9~R_{\rm SNR}$ from the shock front
by this orders-of-magnitude estimation,
which is consistent with our results for Tycho's SNR.
\par
Our results for the inner circles ($R\la0.9~R_{\rm SNR}$),
which show somewhat flatter spectra than the outer ($R\ga0.9~R_{\rm SNR}$) ones at relatively large scales $\lambda/R\ga0.2$,
may be ascribed
to the interaction between the well-developed GS95 turbulence and the RTI driven turbulence.
To examine this interpretation, MHD simulations solving the interaction between the well-developed GS95 turbulence
and the RTI driven turbulence are required.
We will study this in forthcoming paper.
\par
There are other possibilities to amplify the magnetic field at the vicinity of the shock
such as the Bell instability~\citep{bell04},
which is usually expected as the mechanism responsible for the magnetic field amplification
at the {\it upstream} region leading to the acceleration of knee-energy CR protons.
This instability occurs resulting from the interaction between leaking CRs from the shock and the background plasma.
If the upstream field has been already amplified, the trans-Alfv\'enic condition can be satisfied in the region just behind the shock.
This situation may be consistent with our results for Tycho's SNR.
Moreover, the leaking CRs excite an acoustic instability~\citep[enhance a compressible perturbation, e.g.][]{drury86}.
The upstream plasma affected by the leaking CRs, i.e. shock precursor,
interacts with the density fluctuations pre-existing in the ISM.
This interaction also
leads to the field amplification by a turbulent dynamo process
in the upstream region~\citep[][]{beresnyak09,delvalle16}.
Indeed, \citet{xu17} pointed out that
the shock crossing time of the precursor length
is large enough to lead to full development
of the precursor dynamo in partially ionized ISM.
Note that the Balmer line emissions from Tycho's SNR
indicate the interaction between the shock and the partially ionized ISM
~\citep[e.g.][]{chevalier80,lee07}.
\par
The above possible amplification mechanisms predict different evolution tracks
of the magnetic energy spectrum.
For the RMI inducing amplification on the downside
(i.e. turbulent dynamo in super-Alfv\'enic turbulence),
magnetic field disturbances on the scales larger than $l_{\rm A}$,
at which the turbulent velocity is equal to the Alfv\'en velocity,
grow with time~\citep[e.g.][]{cho00a,brandenburg05,xu16}.
It indicates that the $l_{\rm A}$
evolves toward a larger scale with increasing a distance from the shock front.
On the other hand, if the field has significantly been amplified upstream, such evolution would not be seen.
Moreover,
\citet{pohl05} pointed out that the field amplified by the Bell instability damps downstream.
Therefore, we can assess the real amplification mechanism from the spatial variation
of the magnetic energy spectrum.
The correlation for the most outer circle with the radius $R=1.00R_{\rm SNR}$
roughly shows the single power-law with the Kolmogorov-like scaling,
which favors an well-amplified field just behind the shock, i.e. the upstream field amplification.
It would be confirmed once the evolution track of spectrum within the most outer circle is resolved.
\par
Supposing that the strength of amplified magnetic field is $\sim100~{\rm \mu G}$~\citep[e.g.][]{parizot06},
the scale length of the gyroradius of the knee-energy CRs,
$r_{g,{\rm knee}}\sim0.01~{\rm pc}\left(\frac{E}{10^{15.5}~{\rm eV}}\right)\left(\frac{B}{100~{\rm \mu G}}\right)^{-1}$,
is shorter than the spatial resolution of the present data.
The scale length of $r_{g,{\rm knee}}$ will be resolved and the mechanism of field amplification may be distinguished,
if we have data of higher spatial resolution (say, $\la 0.001R_{\rm SNR}$).
Once we obtain the magnetic energy spectrum at the length scale less than $r_{g,{\rm knee}}$,
we can estimate the diffusion coefficient of the knee-energy CRs and
the possibilities of their acceleration in SNR.
We thus need a higher sensitivity with sub arcsecond resolution at GHz band;
this would be a science case of the Square Kilometre Array (SKA).
\par
Finally, our method would be available not only for SNRs
but also other astrophysical objects with a spherical-shell structure,
such as radio relics in galaxy clusters.

\section*{Acknowledgements}

%The Acknowledgements section is not numbered. Here you can thank helpful
%colleagues, acknowledge funding agencies, telescopes and facilities used etc.

We are grateful Dr. Brian J. Williams who kindly provided us with
the VLA image used in this paper.
We thank Prof. Jungyeon Cho for valuable comments to complete this work.
We also thank the anonymous referee for his/her comments
to further improve the paper.
This work is supported by Grant-in-aids
for JSPS Fellows (15J08894, JS) and JSPS KAKENHI Grants: 15H03639,
15K17614, 17H0110 (TA), 15K05039 (TI), and 15K05080 (YF).

%%%%%%%%%%%%%%%%%%%%%%%%%%%%%%%%%%%%%%%%%%%%%%%%%%



\bibliographystyle{mnras}
\bibliography{mnras_salif}


%%%%%%%%%%%%%%%%%%%%%%%%%%%%%%%%%%%%%%%%%%%%%%%%%%

%%%%%%%%%%%%%%%%% APPENDICES %%%%%%%%%%%%%%%%%%%%%

\appendix



\section{Measuring Method of Magnetic Field Correlation}
\label{sec measurements method}
Here we provide the measuring method of magnetic energy spectrum in SNRs
by applying the method developed for the ISM.
\par
\citet{lazarian12,lazarian16} provided a mathematical formalism describing
how second order correlation functions of synchrotron intensities are related with magnetic field disturbances.
An emissivity of a synchrotron emission per frequency $\nu$ depends on the strength of magnetic field
component perpendicular to the line of sight, $|\bm{B}_n|=B_n$, as
%
%%%%%%%%%
\begin{eqnarray}
i_\nu(\bm{r})
&=&K\nu^{-\alpha}
B_n(\bm{r})^{1+\alpha}, \nonumber \\
&=&K\nu^{1-\gamma}
B_n(\bm{r})^{\gamma},
\end{eqnarray}
%%%%%%%%%
%
where $K$ is a function depending on the density of relativistic electrons,
$\alpha=(s-1)/2$, $s$ is the power-law index of the CR electron energy spectrum
and we have defined $\gamma\equiv1+\alpha$ for simplicity.
Thus, the second-order correlation function of the synchrotron emissivity,
%
%%%%%%%%%
\begin{eqnarray}
C^{(2)}_{i_\nu,\gamma}(\bm{l})
&=& \langle i_\nu(\bm{r})i_\nu(\bm{r}+\bm{l})  \rangle_{\bm{r}},
\label{cf I}
\end{eqnarray}
%%%%%%%%%
%
is related with the magnetic field correlation as
%
%%%%%%%%%
\begin{eqnarray}
C^{(2)}_{i_\nu,\gamma}(\bm{l})\propto
\langle B_n(\bm{r}){}^{\gamma}B_n(\bm{r}+\bm{l}){}^{\gamma}  \rangle_{\bm{r}}.
\end{eqnarray}
%%%%%%%%%
%
When $\gamma=1$, which is equivalent to $\alpha=0$ and $s=1$,
and spatial distribution of CR electrons is uniform~(i.e. $K$ is constant),
$C^{(2)}_{i_\nu,\gamma}$ becomes identical to the magnetic field correlation function.
The case of $\gamma=2$ ($\alpha=1$ and $s=3$) is also simple and can be representative
for the case of arbitrary $\gamma$.
Omitting the notations as
$B_n(\bm r)\rightarrow B_n$ and $B_n(\bm{r}+\bm{l})\rightarrow B_n'$,
we obtain
%
%%%%%%%%%%%
\begin{eqnarray}
\langle B_n{}^2 B_n'{}^2 \rangle_{\bm{r}} 
=\frac{\langle B_n{}^4+B_n'{}^4\rangle_{\bm{r}}}{2}
-\frac{1}{2}\left\langle
\left( B_n+B_n' \right)^2
\left( B_n-B_n' \right)^2
\right\rangle_{\bm{r}}.
\label{cf B^2}
\end{eqnarray}
%%%%%%%%%%%
%
If we decompose the magnetic field into the mean component $\bar{B}_n=\langle B_n \rangle_{\bm{r}}$
and the fluctuating component $\Delta B_n=B_n-\bar{B}_n$, the above equation can be written as
%
%%%%%%%%%%%
\begin{eqnarray}
\langle B_n{}^2 B_n'{}^2 \rangle_{\bm{r}} 
&=&\frac{\langle B_n{}^4+B_n'{}^4\rangle_{\bm{r}}}{2} \nonumber \\
&-&2\bar{B}_n{}^4\left\langle
\left( 1+\frac{\Delta B_n+\Delta B_n'}{\bar{B}_n} \right)^2
\left( \frac{\Delta B_n- \Delta B_n'}{\bar{B}_n} \right)^2
\right\rangle_{\bm{r}}. \nonumber \\
\label{cf B^2}
\end{eqnarray}
%%%%%%%%%%%
%
For small standard deviation of the field
$\sqrt{\langle (B_{n} -\bar{B}_n)^2\rangle_{\bm{r}}}/\bar{B}_n\sim
|(\Delta B_n+\Delta B_n')/\bar{B}_n|<1$,
the correlation function of $B_n{}^2$ becomes
%
%%%%%%%%%%%
\begin{eqnarray}
\langle B_n{}^2 B_n'{}^2 \rangle_{\bm{r}} 
&\approx&\frac{\langle B_n{}^4+B_n'{}^4\rangle_{\bm{r}}}{2}
-2\bar{B}_n{}^4 \left\langle
\left( \frac{\Delta B_n- \Delta B_n'}{\bar{B}_n} \right)^2
\right\rangle_{\bm{r}} \nonumber \\
&=& 4\bar{B}_n{}^2\langle \Delta B_n \Delta B_n' \rangle_{\bm{r}}+{\rm const.},
\label{cf B^2}
\end{eqnarray}
%%%%%%%%%%%
%
where we have assumed the isotropic turbulent field as
$\langle B_n{}^4 \rangle_{\bm{r}}=\langle B_n'{}^4 \rangle_{\bm{r}}$ and
$\langle \Delta B_n{}^2 \rangle_{\bm{r}}=\langle \Delta B_n'{}^2 \rangle_{\bm{r}}$.
Thus, $\langle B_n{}^2 B_n'{}^2 \rangle_{\bm{r}}$
reproduces the second-order correlation function of the magnetic field disturbances.
Note that even if we consider a completely random field,
this approximation would be applicable for small scales.
This is because
the field disturbances on the larger scales act as a guide field for
the field disturbances on the smaller scales~\citep[see, e.g.][]{cho00b}. 
\citet{lazarian12} showed that
%
%%%%%%
\begin{eqnarray}
\frac{\langle B_n{}^2 B_n'{}^2 \rangle_{\bm{r}}}
{\langle B_n{}^4 \rangle_{\bm{r}}
- \langle B_n{}^2 \rangle_{\bm{r}}{}^2}
&\approx&
\frac{ \langle B_n{}^\gamma B_n'{}^\gamma \rangle_{\bm{r}} }
{\langle B_n{}^{2\gamma} \rangle_{\bm{r}}
-\langle B_n{}^{\gamma}\rangle_{\bm{r}}{}^2 },
\label{lp12}
\end{eqnarray}
%%%%%%
%
for several $\gamma$ for a power-law correlation function of $B_n$.
In the range of $1.2\le\gamma\le3$, they reported that the maximum difference of $\langle B_n{}^\gamma B_n'{}^\gamma \rangle_{\bm{r}}$
from $\langle B_n{}^2 B_n'{}^2 \rangle_{\bm{r}}$
is only $3\%$.
This suggests that the correlation functions can be written as
%
%%%%%%
\begin{eqnarray}
\langle B_n(\bm{r})^\gamma B_n(\bm{r}+\bm{l})^\gamma \rangle_{\bm{r}}
\approx {\cal P}(\gamma)\langle B_n(\bm{r})^2B_n(\bm{r}+\bm{l})^2\rangle_{\bm{r}},
\end{eqnarray}
%%%%%%
%
where ${\cal P}$ is a function of $\gamma$.
This argument is numerically confirmed by \citet{lee16}.
They performed synthetic observations of synchrotron emissions from simulated magnetic field
and derived the Fourier power spectrum from the correlation of the observed synchrotron polarization intensity
for the parameter range $1.5\le\gamma\le4$.
They found that the power spectrum reproduced
the spectral index of the given magnetic field.
For SNRs, observations of the radio synchrotron intensity per frequency show
the power-law spectrum with the index $\alpha\approx0.6$
that indicates $\gamma\approx1.6$ and $s\approx2.2$
\citep[e.g.][]{green09}.
Hence, $C_{i,\gamma}^{(2)}$ can reproduce
the second-order correlation function of
magnetic field disturbance in the SNRs.
\par
In order to measure the second-order correlation function of
magnetic field disturbance from
the synchrotron emissions, we must consider the projection effect.
We define the observed intensity of synchrotron emission per frequency
at the two-dimensional sky position $\bm{X}=(x,y)$ as
%
%%%%%%%%%
\begin{eqnarray}
I_\nu(\bm{X})=\int_0^{L(\bm{X})} K\nu^{1-\gamma}
B_n(\bm{X},z)^{\gamma}
dz,
\end{eqnarray}
%%%%%%%%%
%
where $z$ represents the coordinate along the line of sight and
$L(\bm{X})$ is the extent of the emission region.
Note that $L$ is a function of $\bm{X}$ in general.
The second-order correlation function for $I_{\nu}$ is written as
%
%%%%%%%%%
\begin{eqnarray}
C^{(2)}_{I_\nu}(\bm{\lambda})
&=&\frac{\int  I_\nu(\bm{X}) I_\nu(\bm{X}') d^2\bm{X}}{ \int d^2\bm{X} }
\\ \nonumber
&\equiv& \langle I_\nu(\bm{X})I_\nu(\bm{X}+\bm{\lambda})  \rangle_{\bm{X}},
\label{cf I}
\end{eqnarray}
%%%%%%%%%
%
where $\bm{\lambda}\equiv\bm{X}'-\bm{X}$ is the position vector of two separated positions in the sky $\bm{X}$ and
$\bm{X}'$~(see also Eq.~\eqref{correlation} of the main text).
The correlation function can be represented as
%
%%%%%%%%%
\begin{eqnarray}
C^{(2)}_{I_\nu}(\bm{\lambda})=K\nu^{1-\gamma}\int_0^{L(\bm{X})} dz \int_0^{L'(\bm{X}')} dz'
\nonumber \\
\times\langle B_n(\bm{X},z){}^{\gamma}B_n(\bm{X}+\bm{\lambda},z'){}^{\gamma}  \rangle_{\bm{X}}.
\nonumber \\
\label{cf I2}
\end{eqnarray}
%%%%%%%%%
%
For the constant $L(\bm{X})=L_0$, \citet{lazarian16} and \citet{lee16} demonstrated
that $C^{(2)}_{I_\nu}$ reproduces the scaling relation of
a given magnetic field correlation.
However, if $L$ varies spatially, $C^{(2)}_{I_\nu}$ is affected by the geometrical structure of
the emission region, which is usually unknown.
However, fortunately, the emission regions of some young SNRs are known to be a spherical shell~\citep[e.g.][]{dickel91,reynoso13}.
Therefore, if we select the points $\bm{X}$ and $\bm{X}'$ on the concentric circle of SNR image,
the condition of $L(\bm{X})=$~constant is satisfied.
%figure geo
%
%%%%%%%%%%%%%%%%%%%%
\begin{figure}
\center
\includegraphics[scale=0.35]{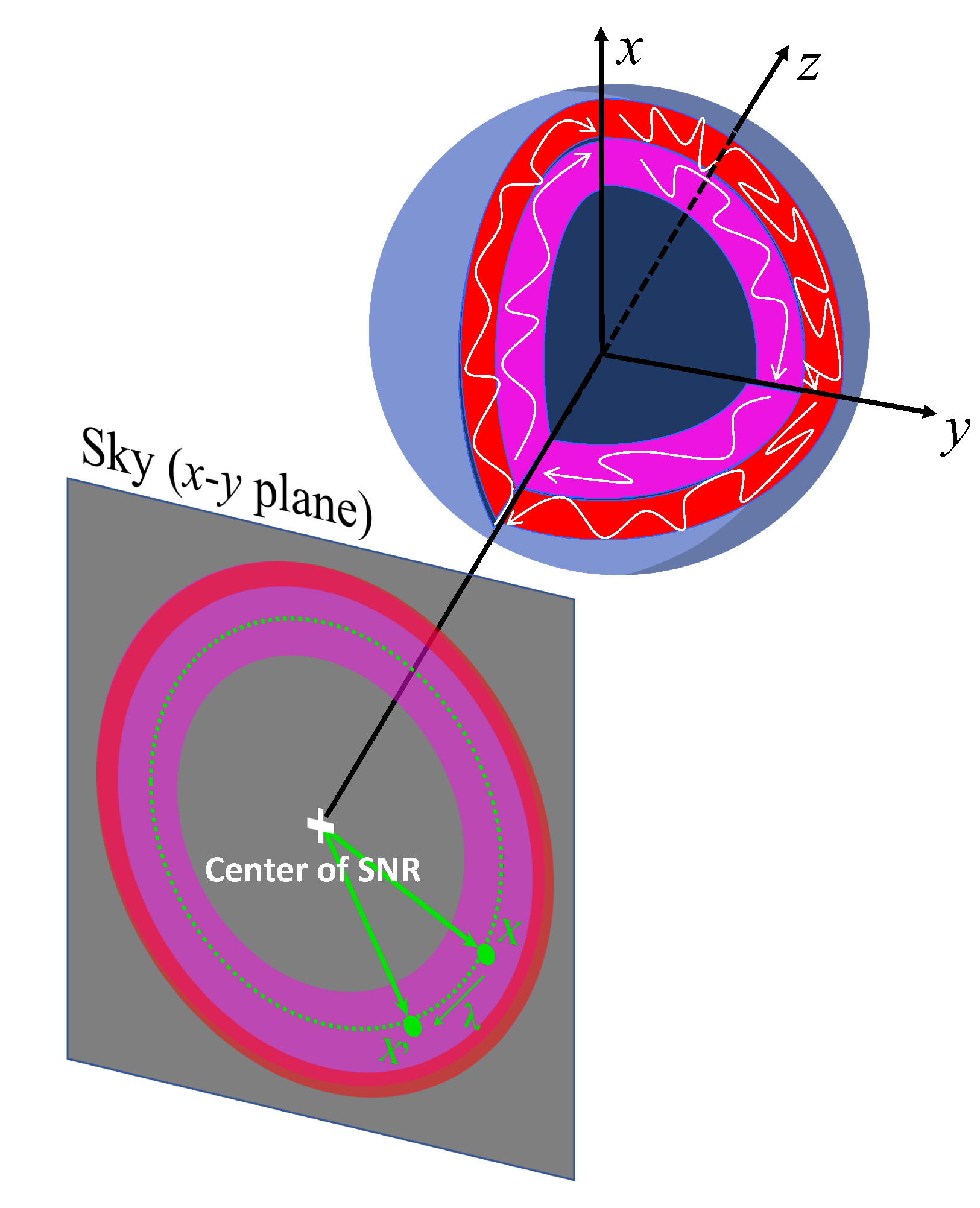}
\caption{Schematic of SNR shell and projected image.
The blue spherical shell with partial cross sections shows the SNR shell.
The line of sight is along the $z$-axis, and the $x\mathchar`-y$ plane corresponds to the projected sky.
The white lines schematically represent turbulent magnetic field lines.
The synchrotron emissions from turbulent media are projected onto the sky (the red and magenta toruses).
The white cross indicates the center of the SNR.
If we analyse the intensity correlation between positions $\bm{X}$ and $\bm{X}'=\bm{X}+\bm{\lambda}$
on the concentric circle (the green dots), the line of sight extent $L(\bm{X})$ becomes
constant and
the correlation function is not affected by the structure of SNR shell.}
\label{fig geo}
\end{figure}
%%%%%%%%%%%%%%%%%%%%
%
%figure geo
In Figure~\ref{fig geo}, we show a schematic of the SNR shell and projected image.

%%%%%%%%%%%%%%%%%%%%%%%%%%%%%%%%%%%%%%%%%%%%%%%%%%

% Don't change these lines
\bsp	% typesetting comment
\label{lastpage}
\end{document}